\documentclass[a4paper,11pt]{article}

\usepackage[english]{babel}
\usepackage[a4paper,tmargin=3.5truecm,bmargin=4truecm,rmargin=2.5truecm,
lmargin=3.2truecm,twoside,verbose=true]{geometry}
\usepackage{cancel,graphicx}
\usepackage{hyperref}
\usepackage{amsmath,amssymb,slashed}

\usepackage{amsmath,amssymb}
\usepackage[amsmath, hyperref, thmmarks]{ntheorem}
\numberwithin{equation}{section}

\allowdisplaybreaks[1]
\newcommand\bbone{{\mathbb{I}}}

\renewenvironment{thebibliography}[1]
         {\section*{References}\frenchspacing\small
          \begin{list}{[\arabic{enumi}]}
         {\usecounter{enumi}\parsep=2pt\topsep 0pt
         \settowidth{\labelwidth}{[#1]}
         \leftmargin=\labelwidth\advance\leftmargin\labelsep
         \rightmargin=0pt\itemsep=1pt\sloppy}}{\end{list}}

\theoremsymbol{}
\theorembodyfont{\slshape}
\theoremheaderfont{\normalfont\bfseries}
\theoremseparator{}

\theorembodyfont{\upshape}
\theoremsymbol{\ensuremath{\blacklozenge}}

\theoremstyle{nonumberplain}
\theoremheaderfont{\scshape}
\theorembodyfont{\normalfont}
\theoremsymbol{\ensuremath{\blacksquare}}

\qedsymbol{\ensuremath{_\blacksquare}}
\theoremclass{LaTeX}

\title{Single Extra Dimension from $\kappa$-Poincar\'e and Gauge Invariance}
\author{Philippe Mathieu$^a$, Jean-Christophe Wallet$^b$}

\begin{document}

\date{}
\maketitle
\vspace*{-1cm}

\begin{center}
\textit{$^a$Department of Mathematics, 
University of Notre Dame, Notre Dame, IN 46556, USA.} \\
\textit{$^b$IJCLab, Universit\'e Paris-Saclay, CNRS/IN2P3, 91405 Orsay, France.}  \\
\textit{}\\
\bigskip
 e-mail:
\texttt{pmathieu@nd.edu, jean-christophe.wallet@th.u-psud.fr}\\[1ex]

\end{center}
\begin{abstract}
We show that $\kappa$-Poincar\'e invariant gauge theories on $\kappa$-Minkowski space with physically acceptable commutative (low energy) limit must be 5-d. The gauge invariance requirement of the action fixes the dimension of the $\kappa$-Minkowski space to $d=5$ and selects the unique twisted differential calculus with which the construction can be achieved. We characterize a BRST symmetry related to the 5-d noncommutative gauge invariance though the definition of a nilpotent operation, which is used to construct a gauge-fixed action. We also consider standard scenarios assuming (compactification of) flat extra dimension, for which the 5-d deformation parameter $\kappa$ can be viewed as the bulk 5-d Planck mass. We study physical properties of the resulting 4-d effective theories. Recent data from collider experiments require $\kappa\gtrsim\mathcal{O}(10^{13})\ \text{GeV}$. The use of standard test of in-vacuo dispersion relations of Gamma Ray Burst photons increases this lower bound by 4 orders of magnitude. The robustness of this bound is discussed in the light of possible new features of noncommutative causal structures. 
\end{abstract}

\vskip 1 true cm
{\it{Dedicated to the memory of John Madore. }}

\vfill\eject
\section{Introduction.}\label{section1}
Noncommutative structures are expected to occur at the Planck scale \cite{dopplic} where Quantum Gravity effects become relevant \cite{amelin}. Among the various noncommutative (quantum) spacetimes, the $\kappa$-Minkowski spacetime \cite{lukie} is believed to be a good candidate to describe the quantum spacetime underlying Quantum Gravity. This noncommutative (quantum) spacetime is known for long to be rigidely linked to the $\kappa$-Poincar\'e algebra \cite{majid-ruegg} coding the quantum version of its relativistic symmetries. This latter already shows up within (2+1)-$d$ gravity with matter as a symmetry of the (effective) Noncommutative Field Theory (NCFT) obtained by integrating out the gravitational degrees of freedom \cite{fredliv} while interesting arguments favoring its role as a symmetry of (3+1)-$d$ quantum gravity (in a particular regime) were given in \cite{AC-ss}, therefore enforcing the belief that at ultra-high energy, Poincar\'e invariance as well as Minkowski spacetime should be replaced by their respective $\kappa$-deformations. The phenomenological consequences \cite{amelin} of these $\kappa$-deformations have been examined in many works, dealing e.g. with Doubly Special Relativity \cite{AC1} or Relative Locality \cite{AC-LS}. \\

For pioneering works on noncommutative gauge theories, see \cite{JMadore}. Since gauge invariance at low energy must supplement Poincar\'e invariance in any reasonable field theory, it is natural to consider noncommutative gauge theories obeying both $\kappa$-Poincar\'e invariance and (noncommutative analog of) gauge invariance at energy near the Planck scale. The quantum properties for NCFT on $\kappa$-Minkowski  spaces stayed rather poorly investigated for a long time \cite{balkan}. Recently, the introduction of a convenient star-product \cite{PW2018}, \cite{DS}{\footnote{For alternative star-products on $\kappa$-Minkowski, see \cite{ru-vit}, \cite{fedele}.}}, resulting from a combination of the Weyl-Wigner quantization with the convolution algebra of the affine group $\mathbb{R}\ltimes\mathbb{R}^{(d-1)}$ ($d$ is the dimension of the $\kappa$-Minkowski space) has allowed us to start studying the one-loop properties of various classes of NCFT on $\kappa$-Minkowski  \cite{PW2018}, \cite{PW2018bis}. \\

But gauge invariance and $\kappa$-Poincar\'e invariance are difficult to reconcile together. In fact, the requirement of $\kappa$-Poincar\'e invariance singles out the simple Lebesgue integral as the trace to be used in the action. But this trace is no longer cyclic w.r.t. the star-product, i.e. it is as a twisted trace. A twist, called the modular twist depending on the dimension $d$ of the $\kappa$-Minkowski space, appears upon cyclic permutation of the factors inside the trace, thus preventing the various factors arising from gauge transformations to balance each other. In \cite{MW20}, we have shown how to combine a twisted noncommutative differential calculus, a twisted notion of noncommutative connection and the modular twist to construct a gauge invariant action quadratic in the curvature in 5 dimensions. This value was shown to be the only one for which the compensations between the twists achieved gauge invariance.\\

The result in \cite{MW20} raised the question to know if it is actually possible to obtain gauge invariance for other values of $d$ (e.g. for $d=4$), by a suitable modification of the twisted differential calculus linked to twisted derivations of the $\kappa$-Poincar\'e algebra and/or the twists related to the noncommutative connection, of course still preserving an acceptable commutative limit for the action. It turns out that the answer is negative as it will be shown in this paper, thus motivating more investigation on physical properties related to the (essentially unique) framework obtained in \cite{MW20}.\\

The paper is organized as follows. In the section 2, we show that $\kappa$-Poincar\'e invariant gauge theories built on the $\kappa$-Minkowski space and constrained to have physically suitable commutative limit must necessarily be 5-d. Gauge invariance of the action can be achieved thanks to the existence of a unique twisted noncommutative differential calculus based on a (unique) family of twisted derivations of the algebra of the deformed translations. 
The subsection \ref{section22} deals with the necessary algebraic constraints on twisted connections, while in the subsection \ref{section23} we show how the gauge invariance requirement fixes the dimension of the $\kappa$-Minkowski space and selects a unique twisted differential calculus. In the section 3, we explore general properties of both 5-d gauge-invariant action possibly coupled to matter and related 4-d effective actions obtained through compactification. The subsection \ref{section31} involves phenomenological aspects devoted in particular to standard scenarios with (compactification of) flat extra dimension \cite{UED} in which the 5-d deformation parameter $\kappa$ is naturally interpreted as the bulk 5-d Planck mass. We study generic properties of the resulting 4-d effective theories, focusing mainly on the zero modes sector. Consistency with recent data from collider experiments \cite{bounds-hep} yields the following lower bound $\kappa\gtrsim\mathcal{O}(10^{13})\ \text{GeV}$. The explicit construction of gauge invariant actions is presented in the subsection \ref{section32}. In the subsection \ref{section33}, we obtain a BRST symmetry linked to the 5-d noncommutative gauge invariance algebraically characterized by a nilpotent operation used to construct a gauge-fixed action. In the section \ref{section34}, we apply the  popular test of in-vacuo dispersion relations of Gamma Ray Burst (GRB) photons \cite{fermitel}, \cite{ESA1}-\cite{ESA3} to the deformed dispersion relation derived from the action and show that the lower bound on $\kappa$ is increased by 4-5 orders of magnitude, favoring the scenarios with very small flat extra dimension. We discuss the robustness of this latter bound in the light of possible new features of noncommutative causal structures. In the section \ref{section4}, we conclude.\\

\section{$\kappa$-Poincar\'e invariant gauge theories on $\kappa$-Minkowski space}\label{section2}
\subsection{Basics properties.}\label{section21}
We use the bicrossproduct basis \cite{majid-ruegg}. Our convention are as in \cite{PW2018}. The relevant formulas are collected in the Appendix \ref{apendixA}. The $d$-dimensional $\kappa$-Minkowski space $\mathcal{M}^d_\kappa$ is conveniently described as the algebra of smooth functions on $\mathbb{R}^d$ with polynomial maximal growth, equipped with the star-product and involution \cite{DS,PW2018} 
\begin{equation}
(f\star g)(x)=\int \frac{dp^0}{2\pi}dy_0\ e^{-iy_0p^0}f(x_0+y_0,\vec{x})g(x_0,e^{-p^0/\kappa}\vec{x})  \label{starpro-4d},
\end{equation}
\begin{equation}
f^\dag(x)= \int \frac{dp^0}{2\pi}dy_0\ e^{-iy_0p^0}{\bar{f}}(x_0+y_0,e^{-p^0/\kappa}\vec{x})\label{invol-4d}.
\end{equation}
Eq. \eqref{starpro-4d} yields 
\begin{equation}
[x_0,x_i]=\frac{i}{\kappa}x_i,\ \ [x_i,x_j]=0
\end{equation}
($[f,g]=f\star g-g\star f$) with $ i,j=1,\cdots,d-1$ describing the usual commutation relations for the $d$-dimensional $\kappa$-Minkowski space. The deformation parameter $\kappa$ has the dimension of a mass and can be naturally identified with the $d$-dimensional Planck mass, not necessarily of the same order of magnitude than the observed 4-d Planck mass $M_P\approx10^{19}\text{GeV}$.\\

The $\kappa$-deformed relativistic symmetries of $\mathcal{M}^d_\kappa$ are coded by the $\kappa$-Poincar\'e algebra $\mathcal{P}_\kappa^d$. Recall that any action $S=\int d^dx\ \mathcal{L}$ where $\mathcal{L}$ is some Lagrangian and $\int d^dx$ is the usual Lebesgue integral is $\kappa$-Poincar\'e invariant. Indeed, one can show \cite{DS} that any element
$h$ in $\mathcal{P}_\kappa^d$ acts on $S$ as 
\begin{equation}
h\triangleright S:=\int d^dx\ h\triangleright\mathcal{L}(\phi)=\epsilon(h)S\label{decadix7777}
\end{equation}
where $\epsilon: \mathcal{P}_\kappa\to\mathbb{C}$ is the counit of $\mathcal{P}_\kappa^d$, the symbol $\triangleright$ denotes the action of $h$ and $\phi$ denotes generically some fields. For instance by using the Appendix \ref{apendixA}, a standard calculation yields 
\begin{equation}
(\mathcal{E}\triangleright \phi)(x)=\phi(x_0+\frac{i}{\kappa},\vec{x})
\end{equation}
where we used
\begin {equation}
\mathcal{E}=e^{-P_0/\kappa}
\end{equation}
and 
\begin{equation}
(P_\mu\triangleright \phi)(x)=-i\partial_\mu \phi(x),\ \ \ \mu=0, \hdots, d-1
\end{equation}
which combined with $\epsilon(\mathcal{E})=1$ and $\epsilon(P_\mu)=0$ leads to $\mathcal{E}\triangleright S=S$, $P_\mu\triangleright S=0$.\\

It is known that the Lebesgue integral satisfies
\begin{equation}
\int d^dx\ (f\star g)(x)=\int d^dx\ \left((\sigma\triangleright g)\star f\right)(x),\label{twistrace}
\end{equation}
\begin{equation}
\sigma=\mathcal{E}^{d-1}\label{twist-pratik},
\end{equation}
i.e. $\int d^dx$ is a twisted trace with respect to \eqref{starpro-4d}. This trades the usual cyclicity for a KMS property. Indeed, as pointed out in \cite{PW2018,matas}, the action $S$ defines a KMS weight \cite{kuster} associated with the (Tomita) group of modular automorphisms \cite{takesaki} 
whose generator is \eqref{twist-pratik}, called the modular twist. For general discussions on physical consequences of KMS property, see \cite{ConRove}. One-loop properties of $\kappa$-Poincar\'e invariant scalar NCFT  on $\mathcal{M}^4_\kappa$ have been examined in \cite{PW2018bis}, showing soft UV behavior, absence of UV/IR mixing and for some of them vanishing of the beta functions \cite{PW2018bis}. \\

Had we decided to abandon the $\kappa$-Poincar\'e invariance, then we could have used a cyclic integral w.r.t. the star product, as in e.g. \cite{arzano}. But, the resulting actions would have had physically unsuitable commutative limits. Notice that the loss of cyclicity does not complicate practical calculations. In fact, any $\mathcal{P}_\kappa^d$-invariant action based on \eqref{starpro-4d}-\eqref{twist-pratik} can be easily represented as a nonlocal field theory involving ordinary integral and commutative product.\\

We will consider the noncommutative analog of $U(1)$ gauge symmetry \cite{mdv}, \cite{cawa}. Generalization to larger symmetry follows from a mere adaptation of \cite{cawa} and would not alter the conclusions obtained in this work. We look for  noncommutative gauge theories on $\mathcal{M}^d_\kappa$ with polynomial actions depending on the curvature (field strength) of the noncommutative connection (gauge potential), to be characterized below, satisfying two requirements:
\begin{itemize}
\item  i) the action, say $S$, is both invariant under $\mathcal{P}_\kappa^d$ and the  noncommutative $U(1)$ gauge symmetry, 
\item ii) the commutative limit of $S$ is physically acceptable, that is it coincides with the action describing an ordinary field theory. 
\end{itemize}

In \cite{MW20}, we have shown that the twisted trace \eqref{twistrace} insuring $\mathcal{P}_\kappa^d$-invariance restricts the allowed values of $d$ at which such an action may eventually exist. One necessary ingredient is the existence of (at least one) suitable twisted  noncommutative differential calculus, the twist being essential. In particular, there is no untwisted differential calculus which can support a gauge invariant action, whatever the dimension of $\mathcal{M}_\kappa^d$ may be \cite{MW20}, as e.g. the bicovariant differential calculi on $\kappa$-Minkowski \cite{sit}. \\

The second ingredient related to the  noncommutative differential calculus is the construction of a twisted connection and its curvature. Requiring the gauge invariance of the action then amounts to require that the effects of the various twists balance the one of the modular twist \eqref{twist-pratik}, resulting in a $d$-depending consistency relation between all the twists. \\
We now show that gauge invariant actions satisfying i) and ii) can only be obtained from a {\it{unique}} 1-parameter family of twisted derivations of the algebra of the ``deformed translations'' $\mathcal{T}^d_\kappa\subset\mathcal{P}_\kappa^d$ (see Appendix \ref{apendixA}) and only for $d=5$, the unique value for which the  noncommutative gauge symmetry can be accommodated with the $\kappa$-Poincar\'e invariance. \\

\subsection{Fixing the twists.}\label{section22}
In this section, there is {\it{no summation over the repeated indices in the formulas}}. \\

We introduce a set of $d$ mutually commuting bitwisted derivations of $\mathcal{M}^d_\kappa$, $\{ X_\mu \}_{\mu=0, \hdots, d-1}$. Recall that $X_\mu$ as a bitwisted derivation \cite{MW20} of $\mathcal{M}_\kappa^d$ is an element of $\mathcal{P}_\kappa^d$ satisfying the twisted Leibniz rule:
\begin{equation}
 X_{\mu}\left(a\star b\right) = X_{\mu}\left(a\right)\star\alpha_\mu\left(b\right)+\beta_\mu\left(a\right)\star X_{\mu}\left(b\right),\label{Leib}
\end{equation}
with 
\begin{equation}
[\alpha_\mu,X_{\mu}]=[\beta_\mu,X_{\mu}]=0.
\end{equation}
 The twists $\alpha_\mu$ and $\beta_\mu$ belong to $\mathcal{P}_\kappa^d$ and are algebra automorphisms of $\mathcal{M}^d_\kappa$, so that
\begin{equation}
\alpha_\mu(a\star b)=\alpha_\mu(a)\star \alpha_\mu(b)
\end{equation}
and the same holds for $\beta_\mu$. Hence, to each $X_\mu$ corresponds a pair of twists $(\alpha_\mu,\beta_\mu)$. The general framework of  noncommutative differential calculi based on such twisted derivations has been characterized in \cite{MW20}. Here, it will be sufficient to work with the ``components'' of the 1-form connection and 2-form curvature. \\

We start from the general bitwisted connection, defined as a map $\nabla_\mu:\mathbb{M}\to\mathbb{M}$, where $\mathbb{M}$ is a module over the algebra, assumed here to be a copy of $\mathcal{M}^d_\kappa$, namely  $\mathbb{M}\simeq\mathcal{M}^d_\kappa$, and satisfying \cite{MW20}
\begin{equation}
\nabla_{\mu}\left(ma\right) = \nabla_{\mu}\left(m\right)\star\tau_\mu\left(a\right)+\rho_\mu\left(m\right)\star X_{\mu}\left(a\right)  
\end{equation}
where $\tau_\mu$ and $\rho_\mu$ are automorphisms of $\mathcal{M}^d_\kappa$, elements of $\mathcal{P}_\kappa^d$. From this follows
\begin{equation}
\nabla_{\mu}(a)=A_{\mu}\star\tau_{\mu}(a)+X_{\mu}(a),\ \ A_{\mu}:=\nabla_{\mu}(1),\label{cnabla}
\end{equation}
where $A_\mu$ is the  noncommutative gauge potential, which will be relevant in the following analysis. A general algebraic presentation of  noncommutative connections (and of derivation-based differential calculi) can be found in \cite{mdv}. For adaptation to the NCFT framework as well as some extensions of the notion of connection, see \cite{cawa}, \cite{jur-wall}. \\

Consider now the most general twisted gauge transformations for which each component of the  noncommutative gauge potential $A_\mu$ is acted on by a left and a right twist, denoted by $\rho_{1,\mu}$ and $\rho_{2,\mu}$. The twisted gauge transformations can be expressed as
\begin{equation}
\nabla_{\mu}(.)
\longrightarrow 
\nabla^{g}_{\mu}(.)
= \rho_{1,\mu}(g^{\dag})\star\nabla_{\mu}\left(\rho_{2,\mu}\left(g\right)\star\cdot\right) \label{gaugetwisted}
\end{equation}
where $\rho_{1,\mu}$ and $\rho_{2,\mu}$ are elements of $\mathcal{P}_\kappa^d$ and are assumed to act as {\it{regular automorphisms}} \cite{como-1} of $\mathcal{M}^d_\kappa$ which implies
\begin{equation}
 \rho_{a,\mu}(g)^\dag=\rho_{a,\mu}^{-1}(g^\dag),\ \ a=1,2,\label{regulier}
\end{equation}
for any $g$ in $\mathcal{M}^d_\kappa$ verifying the unitary relation 
\begin{equation}
g^\dag\star g=g\star g^\dag=1. \label{unitar}
\end{equation}
Here, the group of  noncommutative gauge transformations, denoted by $\mathcal{U}(\mathcal{M}^d_\kappa)$, is {\it{the set of unitary elements of $\mathcal{M}^d_\kappa$, the  noncommutative analog of the $U(1)$ gauge symmetry}}.\\

Now from algebraic manipulations, one infers that $\nabla^g_{\mu}(a)=A^g_{\mu}\tau_{\mu}(a)+X_{\mu}(a)$, $\nabla^g_{\mu}$ given by \eqref{gaugetwisted}, defines a connection if the following relations hold true:
\begin{equation}
\alpha_\mu=\tau_\mu,\ \ \rho_{1,\mu}(g^{\dag})\star\beta_{\mu}\rho_{2,\mu}(g)=1\label{A4}
\end{equation}
\begin{equation}
A^{g}_{\mu} = \rho_{1,\mu}(g^{\dag})\star A_{\mu}\star\tau_{\mu}\rho_{2,\mu}(g)
+ \rho_{1,\mu}(g^{\dag})\star X_{\mu}(\rho_{2,\mu}(g)).\label{grossag}
\end{equation}
\\
The general expression of the components $F_{\mu\nu}$ of the curvature is obtained from the expression
\begin{equation}
\nabla_{\mu}\left(K_{\mu\nu}\nabla_{\nu}\left(a\right)\right)-\nabla_{\nu}\left(K_{\nu\mu}\nabla_{\mu}\left(a\right)\right)
=F_{\mu\nu}\star\tau_{\mu}K_{\mu\nu}\tau_{\nu}\left(a\right),    
\end{equation}
where the twist $K_{\mu\nu}$, element of $\mathcal{P}_\kappa^d$, acts as an automorphism of $\mathcal{M}^d_\kappa$. One easily finds that 
\begin{equation}
F_{\mu\nu}=X_{\mu}\left(K_{\mu\nu}\left(A_{\nu}\right)\right)-X_{\nu}\left(K_{\nu\mu}\left(A_{\mu}\right)\right)
+A_{\mu}\star\tau_{\mu} K_{\mu\nu}\left(A_{\nu}\right)-A_{\nu}\star\tau_{\nu} K_{\nu\mu}\left(A_{\mu}\right),\label{grossfor1}
\end{equation}
is a morphism of (twisted) module provided
\begin{align}
\beta_{\mu} K_{\mu\nu} &= \beta_{\nu} K_{\nu\mu} = 1\label{K_beta},\\
\tau_{\mu} K_{\mu\nu}\tau_{\nu} &= \tau_{\nu} K_{\nu\mu}\tau_{\mu}, \\
\tau_{\mu} K_{\mu\nu} X_{\nu} &= X_{\nu} K_{\nu\mu}\tau_{\mu}, \\
X_{\mu} K_{\mu\nu}\tau_{\nu} &= \tau_{\nu} K_{\nu\mu} X_{\mu}, \\
X_{\mu} K_{\mu\nu} X_{\nu} &= X_{\nu} K_{\nu\mu} X_{\mu}.\label{2}
\end{align}
Upon combining \eqref{grossag} and \eqref{grossfor1}, a tedious calculation leads to the twisted gauge transformations for $F_{\mu\nu}$ given by
\begin{equation}
F^{g}_{\mu\nu}=\rho_{1,\mu}(g^\dag)\star F_{\mu\nu}\star\tau_\mu K_{\mu\nu}\tau_\nu\rho_{2,\nu}(g) \label{megatwist}
\end{equation}
provided the following relations hold true:
\begin{align}
&\tau_{\mu}\rho_{2,\mu}\left(g\right)\star\tau_{\mu} K_{\mu\nu}\rho_{1,\nu}(g^{\dag}) 
= 1 \label{Cd_1},\\
&\beta_{\mu} K_{\mu\nu}\rho_{1,\nu}(g^{\dag}) = \beta_{\nu} K_{\nu\mu}\rho_{1,\mu}(g^{\dag}), \label{22} \\
&\tau_{\mu} K_{\mu\nu}\tau_{\nu}\rho_{2,\nu}(g) = \tau_{\nu}K_{\nu\mu}\tau_{\mu}\rho_{2,\mu}\left(g\right) ,\\
&\tau_{\mu} K_{\mu\nu} X_{\nu}\rho_{2,\nu}\left(g\right) 
= X_{\nu} K_{\nu\mu}\tau_{\mu}\rho_{2,\mu}\left(g\right),\label{222}\\
&X_{\mu} K_{\mu\nu}\rho_{1,\nu}(g^{\dag})
= -\rho_{1,\mu}(g^{\dag})\star X_{\mu}\rho_{2,\mu}(g)\star\tau_{\mu} K_{\mu\nu}\rho_{1,\nu}(g^{\dag})\label{Cd_2}.
\end{align}
\\
We now show that the number of twists is severely restricted, due to compatibility conditions between $(\alpha_\mu,\beta_\mu)$, the twists of gauge transformations $(\rho_{1,\mu},\rho_{2,\mu})$ and $K_{\mu\nu}$. These conditions insure the stability of the space of connections under gauge transformations and (twisted) gauge covariance of the curvature.\\

Indeed, by combining \eqref{K_beta}-\eqref{2} with \eqref{Cd_1}-\eqref{Cd_2}, one realizes that the twists $\rho_{1,\mu}$ are all equal and similarly for $\rho_{2,\mu}$. Namely, 
\begin{equation}
\rho_{1,\mu}=\rho_{1},\ \ \rho_{2,\mu}=\rho_{2},
\end{equation}
for any $\mu=0, 1, \hdots, d-1$, where $\rho_1$ and $\rho_2$ will be characterized in a while. Besides, by combining the unitary relation \eqref{unitar} with eq. \eqref{Cd_1}, one obtains
\begin{equation}
\rho_2=K_{\mu\nu}\rho_1.   
\end{equation}
Hence, all the $K_{\mu\nu}$ are equal to some automorphism of $\mathcal{M}_\kappa^d$ $K$, to be characterized below. Namely, one has
\begin{equation}
K_{\mu\nu}=K, 
\end{equation}
so that \eqref{K_beta} implies 
\begin{equation}
\beta_\mu=\beta=K^{-1},
\end{equation}
and \eqref{Cd_1} yields 
\begin{equation}
\tau_{\mu}=\tau.
\end{equation}
Using $\rho_2=K\rho_1$ and differentiating $\rho_2(g^\dag)\rho_2(g)=1$ by $X_\mu$ using \eqref{Leib}, one can check that \eqref{Cd_2} is verified. \\
Summarizing the above analysis, it appears at this stage that only $\beta$, $\alpha=\tau$ and $\rho_2$ remain as independent twists.\\

Now, assume first that $X_\mu$ belongs to $\mathcal{T}_\kappa^d$. Recall that $\mathcal{T}^d_\kappa$ has primitive elements $(\mathcal{E},P_0,P_i)$ with coproduct 
\begin{equation}
\Delta(\mathcal{E}\otimes\mathcal{E})=\mathcal{E}\otimes\mathcal{E},\ \ \ \Delta(P_0)=P_0\otimes\bbone+\bbone\otimes P_0,\ \ \  \Delta(P_i)=P_i\otimes\bbone+\mathcal{E}\otimes P_i.
\end{equation}
But since $\tau$ and $\beta$ are assumed to be automorphisms of $\mathcal{M}_\kappa^d$, their coproduct must be of the form 
\begin {equation}
\Delta(h)=h\otimes h
\end{equation}
for $h=\tau$, $\beta$. Indeed,  since $\mathcal{M}_\kappa^d$ is a module algebra over $\mathcal{P}_\kappa^d$, one must have $h(a\star b)=m_\star\Delta(h)(a\otimes b)=m_\star(h(a)\otimes h(b))=h(a)\star h(b)$ with $m_\star(a\otimes b)=a\star b$. \\
One therefore concludes that $\tau$, $\beta$ and $K$ must be powers of $\mathcal{E}$, owing to the expression for $\Delta(\mathcal{E})$. Thus, they all are regular automorphisms verifying relations similar to \eqref{regulier}. Since $\mathcal{E}$ commutes with all the elements of $\mathcal{P}_\kappa^d$, all the twists $\beta$, $\tau$ and $\rho_2$ are mutually commuting. \\
\subsection{Gauge and $\kappa$-Poincar\'e invariances - Consistency condition.}\label{section23}
{\it{From now on, summation over repeated indices is understood}}.\\

Now, we look for a gauge invariant action of the form
\begin{equation}
S(A)=\int d^dx\ F_{\mu\nu}\star \big[J_{\mu\nu}(F^{\dag}_{\mu\nu})\big]\label{action-test}
\end{equation}
where $J_{\mu\nu}$ is an automorphism of $\mathcal{M}_\kappa^d$ and the Lebesgue measure $\int d^dx$  involved in $S(A)$ \eqref{action-test} insures that the requirement i) is verified.\\

Upon using \eqref{megatwist} together with \eqref{twistrace} and \eqref{twist-pratik}, one easily finds that \eqref{action-test} is invariant under the NC $\mathcal{U}(\mathcal{M}^d_\kappa)$ gauge transformations provided
\begin{eqnarray}
\mathcal{E}^{d-1}J_{\mu\nu}(\rho_{1}(g^{\dag})^{\dag})\star\rho_{1}(g^{\dag}) &= &1 \label{Cd_31}\\
\tau^{2}K\rho_{2}\left(g\right)\star J_{\mu\nu}\left(\left(\tau^{2}K\rho_{2}\left(g\right)\right)^{\dag}\right)&=&1.\label{Cd_32}
\end{eqnarray}
The combination of eq. \eqref{Cd_31} with \eqref{regulier} and $g\star g^\dag=1$ yields 
\begin{equation}
J_{\mu\nu}=J=\mathcal{E}^{1-d}\rho^2_1. \label{jfinal}
\end{equation}
This, combined with \eqref{Cd_32}, owing to the fact that $\tau$, $K$, $\rho_{1,2}$ commute with each other, gives rise to $\tau^4=\mathcal{E}^{1-d}\beta^4$ where we used $K=\beta ^{-1}$, so that
\begin{equation}
\tau=\mathcal{E}^{\frac{1-d}{4}}\beta\label{decadix1}.
\end{equation}
Using the duality between $\mathcal{M}_\kappa^d$ and $\mathcal{T}_\kappa^d$ \cite{majid-ruegg} and the above restrictions on the twists, one infers from \eqref{Leib} that the coproduct of any $X_\mu$ must be of the form 
\begin{equation}
\Delta(X_\mu)=X_\mu\otimes\tau+\beta\otimes X_\mu. 
\end{equation}
But, as an element of $\mathcal{T}_\kappa^d$, $X_\mu$ must be expressible as a finite sum 
\begin{equation}
X_\mu=\sum x_{mnk}\mathcal{E}^mP_i^nP_0^k. 
\end{equation}
Then, the combination of these two constraints fixes the allowed twisted derivations in $\mathcal{T}_\kappa^d$. One finds after some standard computation that the only possibilities for these latter are
\begin{equation}
\mathcal{E}^\gamma(1-\mathcal{E}),\ \ \mathcal{E}^\gamma P_0,\ \ \mathcal{E}^\gamma P_i,
\end{equation}
where $\gamma$ is a real parameter. The respective pair of twists $(\alpha,\beta)$ are easily found to be given by $(\mathcal{E}^\gamma$, $\mathcal{E}^{\gamma+1})$, $(\mathcal{E}^{\gamma}$, $\mathcal{E}^{\gamma})$ and $(\mathcal{E}^{\gamma}$, $\mathcal{E}^{\gamma+1})$.\\

Finally, notice that the use of twisted derivations out of $\mathcal{T}_\kappa^d$ would lead to actions with unusual (physically unsuitable) commutative limits which would not meet requirement ii). \\

To conclude, using \eqref{decadix1} and $\alpha=\tau$, one finds that the only physically admissible solution is given by 
\begin{equation}
\alpha=\mathcal{E}^{\gamma},\ \  \beta=\mathcal{E}^{\gamma+1}. \label{lestwists}
\end{equation}
This, plugged into \eqref{decadix1}, gives finally
\begin{equation}
1=\mathcal{E}^{\frac{5-d}{4}}\label{finall},
\end{equation}
thus singling out 
\begin{equation}
d=5, \label{result}
\end{equation}
which is independent of $\gamma$. This is the unique physical value for the classical dimension at which $\kappa$-Poincar\'e and NC gauge invariance can coexist, selecting in $\mathcal{P}_\kappa^d$ a unique family of twisted derivations of $\mathcal{T}_\kappa$, given by
\begin{equation}
X^{(\gamma)}_0=\kappa\mathcal{E}^\gamma(1-\mathcal{E}),\ \ X^{(\gamma)}_i=\mathcal{E}^\gamma P_i.\label{sigtau-famil-gamma}
\end{equation}

\section{Discussion.}\label{section3}
\subsection{Phenomenological set-up.}\label{section31}

Let us first summarize the above analysis:\\

Requiring the action on the $\kappa$-Minkowski space $\mathcal{M}_\kappa^d$ to be invariant under the quantum analog of its relativistic symmetries modeled by $\mathcal{P}_\kappa^d$ forces to introduce in the action a twisted trace w.r.t. the star product characterizing $\mathcal{M}_\kappa^d$. The additional requirement of gauge invariance is physically needed in any realistic model on $\mathcal{M}_\kappa^d$ as its commutative limit should at least reproduce the Yang-Mills structure of the Standard Model.\\

As we have shown, the compatibility of $\kappa$-Poincar\'e invariance with ``noncommutative'' (quantum) gauge invariance can be achieved only through the use of twisted connections and curvatures together with corresponding twisted gauge transformations, singling out a unique (one-parameter) family of twisted derivations of the algebra of the so-called deformed translations.\\
From this framework, the various twist effects in an action quadratic in the curvature can balance each other only whenever the classical dimension of $\mathcal{M}_\kappa^d$ is $d=5$ which thus, roughly speaking, appears as a kind of consistency condition for which a (noncommutative) gauge symmetry can be accommodated with the quantum relativistic symmetry of the quantum space. \\

This result goes beyond a pure algebraic interest in the context of noncommutative field theories on $\kappa$-Minkowski space. As a matter of fact, it appears to be a strong physical prediction in that it states clearly that $\kappa$-Poincar\'e invariant gauge theories with physically acceptable commutative (low energy) limit must be 5-d. As a byproduct, this result gives a rationale based on symmetry constraints for the introduction of an extra (spatial) dimension. Note that any experimental evidence disfavoring the existence of a single extra dimension would render questionable the physical relevance of $\kappa$-Poincar\'e invariant {\it{gauge}} theories (together with possibly related concepts linked to $\kappa$-deformations of Minkowski space-time).\\

Matching the 5-d bulk models to 4-d effective theories can be done in principle through various scenarios. These latter depend on various assumptions, in particular those done on the nature of the extra dimension, which can be chosen flat or warped, on its size and on the allowed propagations of the fields which may propagate in the bulk or stay confined on a submanifold/brane. Models with (flat) extra dimension or with warped extra dimension of Randall-Sundrom type are reviewed and compared in e.g. \cite{reviewbrane}  from a phenomenological viewpoint. \\

In this paper, we will consider the first type of scenarios, namely those {\it{based on the assumption of the existence of one (flat) extra dimension}}{\footnote{No restriction is imposed on the field propagation: The fields are assumed to propagate in the bulk.}}. The case of warped dimension will be briefly commented at the end of this subsection.\\

One immediate issue to be discussed is whether any typical contribution coming from the noncommutative structure underlying the 5-d bulk theory may show up in the 4-d effective theory as possibly observable effect which may signal the noncommutative origin of the extra dimension. This depends on the order of magnitude of the deformation parameter $\kappa$ which is the natural mass scale occurring in the present framework. Recall than in a 4 dimensional set-up, the corresponding deformation parameter, say $\kappa_4$,  is usually identified with the Planck mass $M_P$,  namely 
\begin{equation}
\kappa_4\sim M_P,
\end{equation}
which from a Quantum Gravity viewpoint may be naturally viewed as the mass scale which should be observed by all independent observers in a regime where the gravity effects become of the same order of magnitude than those of the other interactions. For an exhaustive review of this interpretation within the Doubly Special Relativity using both the $\kappa$-Poincar\'e (Hopf) algebra as well as, see \cite{camelia5D}.\\

Owing to the above discussion, it is natural to interpret $\kappa$ as the 5-d bulk Planck mass. The requirement ii) described in the section \ref {section21} implies that the generic expression for any 4-d effective theory should take the form 
\begin{equation}
S_{4}=S^{(\kappa\to\infty)}_{4}+\Delta S_{4}
\end{equation}
where in obvious notations $\Delta S_4$ involves all the noncommutative corrections of the order $\mathcal{O}(\frac{1}{\kappa^n})$, $n\ge1$, to $S^{(\kappa\to\infty)}_{4}$ which is some usual commutative action independent of $\kappa$. Of course, a similar decomposition holds for the bulk action, namely $S_{5}=S^{(\kappa\to\infty)}_{5}+\Delta S_{5}$. \\

In a wide class of phenomenologically interesting scenarios, underlying the popular models with flat extra dimensions or UED models \cite{UED}, one assumes that the space is flat while the extra dimension is compact with size $R\sim\frac{1}{\mu}$, where $\mu$ is some mass scale, constrained as to be in agreement with experimental data. In these scenarios, the popular relation relating $\mu$, $M_P$ and the (5-d) bulk Planck mass, say $M_\star$, is given by $M_P^2\approx M_\star^3\frac{1}{\mu}$. Recall that it simply stems from a mere 5-d extension of the Einstein-Hilbert action combined with a standard dimensional analysis while it is assumed that it holds true also for gauge theory models \cite{reviewbrane}. \\

We will follow the same line of thought in this paper, which is a natural assumption, so that the above relation between the mass scales becomes
\begin{equation}
 M_P^2\approx\kappa^3\frac{1}{\mu}\label{decadix-bis},
\end{equation} 
since $\kappa$ is identified with the 5-d bulk Planck mass $M_\star$, while the fields are assumed to propagate in the bulk.\\

Recent collider experiments \cite{bounds-hep} have produced lower bounds for the size of the extra dimension $\mu^{-1}$ within flat extra-dimensional and UED scenarios \cite{UED}. From these analysis, one can set conservatively
\begin{equation}
\mu\gtrsim\mathcal{O}(1-5)\ \text{TeV}.     
\end{equation}
This combined with \eqref{decadix-bis} and assuming $M_P\sim\mathcal{O}(10^{19})$ GeV yields
\begin{equation}
\kappa\gtrsim\mathcal{O}(10^{13})\ \text{GeV}.\label{grosscale}
\end{equation}

At this stage, some remarks are in order which will be illustrated in the next subsection within a generic $\mathcal{U}(\mathcal{M}^5_\kappa)$ gauge theory coupled to a fermion.
\begin{enumerate}
\item First, recall that, in the flat extra dimensional scenarios, the reduction from 5 dimensions to 4 dimensions is usually done through a suitable compactification of the extra dimension, for instance on $\mathbb{S}^1/\mathbb{Z}_2$, leading basically to an effective 4-d theory involving ``light'' ($\mu$-independent) zero-modes plus ``heavy'' Kaluza-Klein (K-K)-type modes with masses proportional to $\mu$. \\
Disregarding the K-K sector and focusing first on the zero-mode sector, one can already expect that corrections should be affected by factors $\sim\mathcal {O}(\frac{\left\langle E\right\rangle}{\kappa})$ where $\left\langle E\right\rangle$ is the energy scale at which a given process is observed. In view of the value for the lower bound of $\kappa$ \eqref{grosscale}, this implies in particular a strong suppression of these possible corrections by factors $\sim\mathcal {O}(\frac{\sqrt{s}}{\kappa})$ in present and future collider experiments, where $\sqrt{s}$ is, as usual, the energy in the center of mass frame.\\
In the same way, note that in the K-K sector, the corrections to the K-K number conservation law \cite{UED} stemming from the $\kappa$-deformation (which will be apparent from \eqref{interac}) would lead to unobservable effects at currently reachable energy scale, since these corrections are also suppressed by inverse powers of $\kappa$. 
\item The fact that the set of twisted derivations \eqref{sigtau-famil-gamma} underlies the present framework together with the requirement of $\mathcal{U}(\mathcal{M}^5_\kappa)$  gauge-invariance of the pure gauge action necessarily leads to a ($\kappa$-depending) deformed ``dispersion relation'' for the gauge potential $A_\mu$ in the 5-d theory which gives rise to a deformed dispersion relation for the photon in the 4-d effective theory. Potentially observable effects from deformed dispersion relations have been already discussed in many works. For exhaustive reviews, see e.g. ref. \cite{amelin}. These effects have been discussed either by starting from general postulated parametrizations of the dispersion relation, or within specific classes of quantum gravity models producing dispersion relations consistent with the postulated parametrizations. \\
Among the related cosmological/astrophysical tests currently considered in the literature \cite{amelin}, we will focus in the subsection \ref{section33} on the constraints on the 5-d bulk Planck mass $\kappa$ obtained from GRB data \cite{fermitel}, \cite{ESA1}-\cite{ESA3}.
\item We note that the noncommutative gauge symmetry, i.e. the  gauge symmetry introduced in the subsection \ref{section22}, is broken through the reduction from 5-d to 4-d. This is a mere corollary of the main result derived in this paper. The resulting 4-d effective theory will however involve a part which is invariant under the usual $U(1)$ gauge symmetry plus additional $U(1)$ gauge breaking terms suppressed by powers of $\frac{1}{\kappa}$. These ``small'' terms will of course disappear at the commutative $\kappa\to\infty$ limit, leading to a $U(1)$ invariant action.\\
This can be easily realized for instance by first considering the (5-d) field strength $F_{\mu\nu}$ eqn. \eqref{grossfor1} or \eqref{pratik-f}. This latter expanded up to the 2nd order in $\frac{1}{\kappa}$ yields an expression of the form 
\begin{equation}
F_{\mu\nu}=f_{\mu\nu}+\frac{1}{\kappa}\phi_{\mu\nu}+\mathcal{O}(\frac{1}{\kappa^2}),
\end{equation}
in which $f_{\mu\nu}$ is the usual abelian $U(1)$ field strength while $\phi_{\mu\nu}$ can be read off from \eqref{pratik-f}. Then, by using the action \eqref{action-test} rescaled by $\frac{1}{g_5^2}$ where $g_5$ is the 5-d gauge coupling constant, a standard calculation gives rise to the $\kappa$-expanded action having the generic expression
\begin{equation}
S=\frac{1}{g_5^2}\int d^5x\ (f_{\mu\nu}f_{\mu\nu}+\frac{1}{\kappa}\phi_{\mu\nu}f_{\mu\nu}+\mathcal{O}(\frac{1}{\kappa^2})),
\end{equation}
(summing implicitely over $\mu$ and $\nu$ from $0$ to $4$) where only the first term is invariant under the (5-d) $U(1)$ gauge symmetry. \\
Then, a standard compactification of the extra dimension gives rise to a 4-d action of the form 
\begin{equation}
S_{\mathrm{eff}}=\frac{1}{g^2}\int d^4x\ (f^0_{mn}f^0_{mn}+\frac{1}{\kappa}\Delta(A)+...) 
\end{equation}
where the ellipses denote K-K heavy modes and 
\begin{equation}
g^2=\mu g_5^2\label{g4}
\end{equation}
is the 4-d gauge coupling constant. Again, only the first term is invariant under the (now 4-d) $U(1)$ gauge symmetry with $f^0_{mn}=P_mA^0_n-P_nA^0_m$ ($m,n=0,1,2,3$) being identified with the usual 4-d $U(1)$ field strength, the superscript $^0$ denoting the zero mode part while $\frac{1}{\kappa}\Delta(A)$ collects all the terms which are not $U(1)$ invariant. In this respect, it may be interesting to build a model involving (at least) one family of leptons using the present framework and examine if the terms involves in $\frac{1}{\kappa}\Delta(A)$ may induce possibly observable effects.
\item Finally, one may wonder if the above conclusions on the various mass scales and in particular \eqref{grosscale} linked to the validity of the relation \eqref{decadix-bis} would be possibly modified assuming a scenario with a warped extra dimension \cite{reviewbrane}. A definite conclusion would await for the proper inclusion, if possible at all, of a suitable warped metric into the present framework. This is beyond the scope of this paper. 
\end{enumerate}

\subsection{The gauge invariant action.}\label{section32}

Let us discuss general physical features of $\kappa$-Poincar\'e invariant gauge theories. In this subsection, we consider the coupling of $S(A)$ \eqref{action-test} to a fermion $\psi$, assuming from now on that $A_\mu$ is real-valued and $\rho_2=\bbone$. \\

A family of $\mathcal{U}(\mathcal{M}^d_\kappa)$ gauge invariant actions can be obtained by starting from
\begin{equation}
\nabla_{\mu}(\psi)=A_\mu\star\mathcal{E}^{\gamma}(\psi)+X^{(\gamma)}_{\mu}(\psi),\label{sigtauconter}
\end{equation}
while the curvature can be expressed as
\begin{equation}
F_{\mu\nu}=\mathcal{E}^{-1}(X^{(0)}_\mu A_\nu-X^{(0)}_\nu A_\mu)+A_\mu\star\mathcal{E}^{-1}(A_\nu)-A_\nu\star\mathcal{E}^{-1}(A_\mu)\label{pratik-f}
\end{equation}
where $X^{(0)}_\mu$ is given by \eqref{sigtau-famil-gamma} in which $\gamma=0$.\\

From the analysis performed in the previous section, one easily infers that the gauge transformations are given by
\begin{equation}
\nabla_{\mu}^g(\psi)=\mathcal{E}^{\gamma+1}(g)\star\nabla_{\mu}(g^\dag\star \psi),\label{gaugebitwist1}
\end{equation}
\begin{equation}
A^{g}_{\mu} = \mathcal{E}^{\gamma+1}(g^{\dag})\star A_{\mu}\star\mathcal{E}^{\gamma}(g) 
+\mathcal{E}^{\gamma+1}(g^{\dag}) \star X^{(\gamma)}_{\mu}(g),\label{grossag-bis}
\end{equation}
\begin{equation}
F_{\mu\nu}^g=\mathcal{E}^{\gamma+1}(g)\star F_{\mu\nu}\star\mathcal{E}^{\gamma-1}(g^\dag).\label{gaugebitwist2}
\end{equation}
where we used the fact that the matter gauge transformations are untwisted, i.e. they read $\psi^g=g\star \psi$, as assumed above. Recall that twisted matter gauge transformations are defined by $\psi^g=\rho_2(g)\star\psi$. \\The extension to a twist $\rho_2\ne\bbone$  is straightforward and would not alter the conclusions of this paper.\\

The above material, combined with \eqref{jfinal} and \eqref{lestwists}, so that 
\begin{equation}
J=\mathcal{E}^{2(\gamma-1)}, 
\end{equation}
together with \eqref{action-test} gives rise to the $\mathcal{U}(\mathcal{M}^d_\kappa)$ gauge invariant classical action given by
\begin{equation}
S=S(A)+S(\psi,A)=\int d^5x\ (\frac{1}{2g_5^2}F_{\mu\nu}\star \mathcal{E}^{2(\gamma-1)}(F^{\dag}_{\mu\nu})+\overline{\psi}\star\mathcal{E}^{-\gamma-1}\slashed{\nabla}\psi)\label{model1} 
\end{equation}
with $\slashed{\nabla}=\gamma^\mu\nabla_\mu$, $\overline{\psi}=\psi^\dag\gamma_0$ in obvious notation ($\gamma_0^\dag=\gamma_0$) and the coupling constant $g_5^2$ has mass dimension $-1$. \\

It is convenient to re-express \eqref{model1} by introducing the Hilbert product 
\begin{equation}
\langle a,b \rangle:=\int d^5x\ a^\dag\star b\label{Hilberrt}
\end{equation}
defined in \cite{PW2018}. It yields 
\begin{equation}
S(A)=\frac{1}{2g_5^2}\langle F_{\mu\nu},\mathcal{E}^{-2(\gamma+1)} F_{\mu\nu}\rangle\label{fmunurangle}
\end{equation}
showing that $S(A)$ is real.\\
It can be noticed that the twisted derivations $X_\mu^{(\gamma)}$ are self-adjoint operators. Namely, one has
\begin{equation}
\langle a, X_\mu^{(\gamma)}(b)\rangle=\langle X_\mu^{(\gamma)}(a), b\rangle\label{selfadj}.
\end{equation}
Eqn \eqref{selfadj} amounts to show that the primitive elements of $\mathcal{T}_\kappa$, $\mathcal{E}$ and $P_i$, are self-adjoint. Indeed, one easily realizes that $\langle \mathcal{E}a,b\rangle=\int d^5x\ (\mathcal{E}a)^\dag\star b=\int d^5x\ \mathcal{E}^{-1}a^\dag\star b=\int d^5x\ a^\dag\star\mathcal{E}b=\langle a,\mathcal{E}b\rangle$ where we used successively \eqref{pairing-involution} and \eqref{decadix7777}. \\
In a similar way, one infers $\langle P_ia,b\rangle=\langle a,P_ib\rangle$ stemming from the Leibnitz rule for $P_i$ implied by the second relation in \eqref{hopf1}, namely $P_i(a\star b)=P_i(a)\star b+\mathcal{E}(a)\star P_i(b)$. We assume here that $a$ and $b$ are Schwarz functions, which will be actually the case of the classical fields involved in the action. \\

To obtain a gauge invariant coupling of $\psi$ to the gauge potential $A_\mu$, one looks for a term of the form $\langle\gamma_0\psi ,\mathcal{E}^{x}\nabla_\mu(\psi) \rangle$ where $x$ should be determined to insure the gauge invariance. By making use of \eqref{gaugebitwist1} so that 
\begin{equation}
\nabla^g_{\mu}(\psi^g)=\mathcal{E}^{\gamma+1}(g)\star\nabla_{\mu}(\psi),
\end{equation}
one can write
\begin{eqnarray}
\langle\gamma_0\psi^g ,\mathcal{E}^{x}\nabla^g_\mu(\psi^g) \rangle&=&\langle{\gamma_0}g\star\psi ,\mathcal{E}^{x}\mathcal{E}^{\gamma+1}(g)\star\mathcal{E}^{x}\nabla_\mu(\psi) \rangle\nonumber\\
&=&\langle\gamma_0\psi ,g^\dag\star\mathcal{E}^{x}\mathcal{E}^{\gamma+1}(g)\star\mathcal{E}^{x}\nabla_\mu(\psi) \rangle\label{breuu}
\end{eqnarray}
in which we used $\langle g^\dag\star a,b\rangle=\langle a, g\star b\rangle$ together with \eqref{selfadj} and 
\begin{equation}
(\mathcal{E}(g))^\dag=\mathcal{E}^{-1}(g^\dag)
\end{equation}
to obtain the second equality in \eqref{breuu}. It follows that gauge invariance is achieved provided
\begin{equation}
 x=-\gamma-1. 
\end{equation}
Putting all together, one easily realize that eqn \eqref{model1} can finally be re-expressed as
\begin{equation}
S=S(A)+S(\psi,A)=\frac{1}{2g_5^2}\langle F_{\mu\nu},\mathcal{E}^{-2(\gamma+1)} F_{\mu\nu}\rangle+\langle\gamma_0\psi ,\mathcal{E}^{-(\gamma+1)}\slashed{\nabla}(\psi) \rangle,\label{model1-trans}.
\end{equation}
The above action can be enlarged by the addition of a gauge invariant fermion mass term given by
\begin{equation}
S_m=m\int d^5x\ \overline{\psi}\star\psi=m\int d^5x\ \overline{\psi}\mathcal{E}^4\psi \label{psim}
\end{equation}
where $m>0$. 

The $\kappa\to\infty$ limit of \eqref{model1} and \eqref{model1-trans} obviously yields the usual (5-d) QED action, while the $\mathcal{U}(\mathcal{M}^d_\kappa)$ invariance reduces to the standard $U(1)$ gauge invariance at this limit. \\

To illustrate the first remark given at the end of the subsection \ref{section31}, we consider the gauge-matter interaction. This latter can be written as 
\begin{align}
S_{\mathrm{int}}=\int\prod_{i=1}^3 \frac{d^5k_i}{(2\pi)^5}\slashed{A}(k_1)\psi(k_2)\overline{\psi}(k_3)\delta^{(1)}(k_1^0+k_2^0+k_3^0)
\delta^{(4)}(\vec{k_1}+\vec{k_2}e^{-k_1^0/\kappa}+\vec{k_3})\times e^{\frac{\gamma-3}{\kappa}k_3^0} \label{interac},
\end{align}
where $k_a=(k_a^0,\vec{k}_a)$, $a=1,2,3,4$. Eqn. \eqref{interac} exhibits the usual energy conservation as indicated by the first delta function, while the momentum conservation is ``deformed'' by the factor ${e^{-k_1^0/\kappa}}$ occurring in the second delta function. \\

As far as the effective 4-d theory is concerned, the compactification of the extra dimension, for instance on $\mathbb{S}^1$ or $\mathbb{S}^1/\mathbb{Z}_2$, gives rise to a Kaluza-Klein number conservation law, see for instance\cite{UED}. This latter law however is deformed stemming from the deformation of the momentum conservation resulting in a delta function of the form 
\begin{equation}
\delta({n_1}+{n_2}e^{-k_1^0/\kappa}+n_3),\label{pasglop}
\end{equation}
in which $n_1,n_2,n_3$ are integers (the so called K-K numbers) generated upon the above compactifications by the periodicity on the fifth coordinate (say $x_4$) of the various fields, leading to a discretization of the fifth momentum component. \\

As it can be seen from \eqref{pasglop}, the possible effect of the deformation depends on the magnitude of $\kappa$ controlling  the contribution of the noncommutativity, which however will be strongly suppressed by inverse powers of $\kappa$ due to the lower bound \eqref{grosscale} as already indicated in the first remark of the subsection \ref {section31}. These effects will be unobservable in collider experiments, being suppressed by factors $\sim\mathcal{O}(\frac{k^0}{\kappa})\sim\mathcal{O}(\frac{\sqrt{s}}{\kappa})$. \\A similar conclusion holds true for the $A_\mu$ self-interactions.\\

\subsection{BRST gauge-fixing and kinetic operator.}\label{section33}

We consider the kinetic term for $A_\mu$ in the action \eqref{model1} and \eqref{model1-trans}. By making use in the quadratic part of \eqref{model1} of the useful identity 
\begin{equation}
\int d^5x\ (f\star g^\dag)(x)=\int d^5x\ f(x)\bar{g}(x),
\end{equation}
which holds true for any (Schwartz) functions, together with \eqref{selfadj}, one easily realizes that the kinetic term can be cast into the form
\begin{equation}
S_{\mathrm{kin}}(A)=\frac{1}{g_5^2}\int d^5x\ A_\mu\mathcal{E}^{-2\gamma}({X^{(0)}_{\alpha}}^2\delta_{\mu\nu}- X^{(0)}_\mu X^{(0)}_\nu)A_\nu\label{skin-gauge}. 
\end{equation}
The second term in \eqref{skin-gauge} can be gauged away by using the gauge condition 
\begin{equation}
X^{(0)}_\mu A_\mu=0,
\end{equation}
which reduces to the usual Lorentz gauge condition in the commutative limit. \\

It is convenient to introduce at this stage the BRST symmetry which can be used for
the implementation of the gauge-fixing. This is done by adding to $S(A)$ in \eqref{model1} the gauge-fixing term 
\begin{eqnarray}
S_{\mathrm{GF}}&=&s\int d^5x \  \big({\overline{C}}^\dag\star\mathcal{E}^{-4}(X^{(0)}_\mu A_\mu) \big)\nonumber\\
&=&\int d^5x\ bX^{(0)}_\mu A_\mu-\int d^5x\ {\overline{C}}^\dag\star\mathcal{E}^{-4}(X^{(0)}_\mu s A_\mu),\label{baooo}
\end{eqnarray}
where the BRST operator $s$ verifies 
\begin{equation}
s^2=0
\end{equation}
and is defined here by the following structure equations
\begin{eqnarray}
sA_\mu&=&X^{(\gamma)}_\mu(C)+A_\mu\star \mathcal{E}^\gamma(C)-\mathcal{E}^{\gamma+1}(C)\star A_\mu,\\ 
sC&=&-C\star C,\\
s\overline{C}&=&b,\\
sb&=&0 \label{brs},
\end{eqnarray}
which imply
\begin{equation}
sF_{\mu\nu}=F_{\mu\nu}\star\mathcal{E}^{\gamma-1}(C)-\mathcal{E}^{\gamma+1}(C)\star F_{\mu\nu}\label{sf}.
\end{equation}
Here, the fields $C$, $\overline{C}$ and $b$ are respectively the ghost, antighost and St\"uckelberg auxiliary field assumed to be real valued. Recall that the functional integration over $b$ serves to implement the gauge condition. The respective ghost numbers of $C$, $\overline{C}$ and $b$ are $1$, $-1$ and $0$. One has $ C^\dag=-C $ stemming from \eqref{unitar}. In \eqref{baooo}, the second term involves the kinetic term for the ghosts as well as the ghost-gauge potential interaction which will not be of our concern in the following.\\

A complete algebraic characterization of the BRST symmetry used in the present context, which bears some common features (but also some differences) with the standard BRST symmetry arising in commutative gauge theories \cite{storwal} will be presented elsewhere \cite{PMJCW21}. Here, we recall that $s$ acts as a derivation w.r.t. the grading defined by the sum of the ghost number and the degree of forms modulo 2. In particular, for any functions (0-forms) $a$ and $b$, one has
\begin{equation}
s(a\star b)=s(a)\star b+(-1)^{\delta(a)}a\star s(b),
\end{equation}
where $\delta(a)$ denotes the ghost number of $a$ and $s(a^\dag)=(s(a))^\dag$ together with 
\begin{equation}
[s,X^{(\gamma)}_\mu]=0. 
\end{equation}
\\
As expected, $S(A)$ is BRST invariant, namely 
\begin{equation}
sS(A)=0, 
\end{equation}
as it can be verified upon using \eqref{sf} combined with the expression for $S(A)$ in \eqref{model1}, so that the gauge-fixed action $S(A)+S_{GF}$ is BRST invariant. \\

It can be easily verified that, in the commutative limit, this gauge-fixed action reduces to the usual gauge-fixed QED action in the Lorentz gauge together with the related BRST structure equations. \\

Upon integrating out the $b$ field, the kinetic term reduces to
\begin{equation}
S_{\mathrm{kin}}(A)=\frac{1}{g_5^2}\int d^5x\ A_\mu \mathcal{K}(\kappa)A_\mu
\end{equation}
with 
\begin{equation}
\mathcal{K}(\kappa)=\mathcal{E}^{-2\gamma}(\kappa^2(1-\mathcal{E})^2+\vec{P}^2) ,   \label{laplacien}
\end{equation}
in which we further assume the following relation
\begin{equation}
0<\gamma<1 
\end{equation}
which insures a suitable decay at large momenta for the propagator.\\

It is easy to realize that $\mathcal{K}(\kappa)$ is related to the Casimir operator of $\mathcal{P}^5_\kappa$
\begin{equation}
C^1_\kappa=\mathcal{E}^{-1}(\kappa^2(1-\mathcal{E})^2+\vec{P}^2)\label{casimirr}
\end{equation}
by the relation
\begin{equation}
\mathcal{K}(\kappa)=\mathcal{E}^{-2\gamma+1}C^1_\kappa.\label{kc1}
\end{equation}
\\
We point out that within the present framework, there is almost no freedom in the choice of the $\mathcal{U}(\mathcal{M}^5_\kappa)$ action quadratic in the curvature $S(A)$ in \eqref{model1}. Its expression is strongly constrained by the relevant differential calculus \eqref{sigtau-famil-gamma} related to the distinguished family of twisted derivations of $\mathcal{T}_\kappa$ given in the subsection \ref{section23}, together with gauge invariance and $\kappa$-Poincar\'e invariance. From this follows the expression for the kinetic operator as well as for the deformed dispersion relation that will emerge in the 4-d effective action for the zero modes, to which we turn now on.

\subsection{Deformed dispersion relations and Gamma Ray Bursts.}\label{section34}
It is natural to identify $\mathcal{K}(\kappa)$ to the noncommutative analog of the 5-d Laplacian. In this respect, \eqref{laplacien} or \eqref{kc1} already signal the appearance of a deformed energy-momentum relation in the 4-d effective theory describing the {\it{zero modes}}. These latter are naturally identified here to the photons. \\

We recall that such deformed dispersion relations have been widely considered in the literature to study possible observable effects on the light speed energy dependence for electromagnetic signals traveling on cosmological distances from a distant source to Earth. So far, the corresponding analysis gave rise to plausible estimates for lower bounds on the Quantum Gravity mass scale, say $M_{qg}$, or more precisely to any typical large mass scale, denoted below by $\overline{M}$,  related to the origin of the energy dependence of the light speed, which in a 4-d context, is usually identified either with $M_{qg}$ or with $M_P$. For a review on the huge phenomenological literature on the subject, see e.g. \cite{amelin} and references therein. \\
Gamma Ray Bursts (GRB)  \cite{fermitel} are known to be suitable source candidates. Indeed, the very large distance the emitted energetic photons{\footnote{Observed energy of the emitted photons can reach $\mathcal{O}(100)$ GeV; see e.g. \cite{fermitel}.}} travel, possibly of the order of several billion of light years, can ``amplify'' the effect of the extremely small correction on the photon velocity. This may result into observable quantities, in particular the arrival time lags between photons of different energies \cite{ESA1}-\cite{ESA3}.  \\

Among the above studies, those dealing with the area of $\kappa$-deformations are of our concern here. It appears that most of them do not resort to some Lagrangians from which dispersion relations could be derived from first principles and standard calculations. Instead, they start from {\it{postulated/conjectured}} reasonable parametrizations for deformed dispersion relations depending on a set of parameters involving $\overline{M}$, to be fitted/constrained by the observed data. Of course, (almost all) these parametrizations reduce for a given set of parameters to the dispersion relation one would obtain from the 4-d version of the deformed Laplacian \eqref{casimirr}, \eqref{laplacien}. This phenomenological approach based on in-vacuo dispersion of GRB photons is expected to give rise to indicative bounds on the Quantum Gravity mass scale which are reviewed in \cite{amelin}. \\

Let us first follow this pragmatical route. To adapt this approach to the present situation, one first notice from that the relevant mass scale appearing in the 4-d effective theory for the zero modes (hereafter denoted by $a_\mu$) is $\kappa$, i.e $\overline{M}=\kappa$. The corresponding action takes the generic form
\begin{equation}
S^0_{\mathrm{eff}}(a_\mu)=\int d^4x\ \big(a_\mu K_4(\kappa)a_\mu+g(\frac{1}{\kappa}V_3(a_\mu))+g^2(\frac{1}{\kappa}V_4(a_\mu)) \big)\label{act-eff},
\end{equation}
in which we have rescaled $a_\mu$ as $a_\mu\to ga_\mu$ where $g$ is the 4-d coupling constant given by \eqref{g4}, $K_4(\kappa)$ is simply the natural reduction of $\mathcal{K}$ to 4-d, namely
\begin{equation}
K_4(\kappa)=\mathcal{E}^{-2\gamma}(\kappa^2(1-\mathcal{E})^2+\vec{P}^2),\ \ \ \ \vec{P}=(P_1,P_2,P_3),
\end{equation}
and $V_3(a)$ and $V_4(a)$ are respectively cubic and quartic self-interaction terms among the $a_\mu$ which represent NC corrections. They are at least of the order of $\frac{1}{\kappa}$ and can be obtained by expanding the interaction terms in $S(A)$ \eqref{model1} in powers of $\frac{1}{\kappa}$.\\
Then, neglecting the self-interactions{\footnote{These may enter through 1st order corrections in the 2-point function; the leading contribution is $\mathcal{O}(\frac{g^2}{\kappa})$ which, assuming $g^2\sim\alpha_{em}\sim\mathcal{O}(10^{-2}) $, can be neglected compared to the classical contributions $\mathcal{O}(\frac{1}{\kappa})$ stemming from the Taylor expansion of the first term in \eqref{act-eff}.}}, \eqref{act-eff} yields the equation of motion $K_4(\kappa)a_\mu=0$ so that the corresponding dispersion relation reads
\begin{equation}
\kappa^2(1-e^{-E/\kappa})^2+\vec{p}\ ^2=0 \label{dispersion}
\end{equation}
in which $\vec{p}$ is a 3-momentum and we have set $p_0=E$. \\

Now, expanding \eqref{dispersion} up to $\mathcal{O}(\kappa^{-2})$ (and switching to Minkowskian signature for convenience), one obtains 
\begin{equation}
E^2-|\vec{p}| ^2-\frac{1}{\kappa}E^3+\mathcal{O}(\frac{1}{\kappa^2})=0. 
\end{equation}
This latter expression, confronted to past studies exploiting GRB data to look for light speed variations \cite{ESA1}--\cite{ESA3}, gives rise to an indicative bound on $\kappa$. Putting all the results of  \cite{ESA1}--\cite{ESA3} together, one can be reasonably keep the following value
\begin{equation}
\kappa\gtrsim\mathcal{O}(10^{-2}-10^{-1})M_P.\label{kappagbr}
\end{equation}
Recall by the way that this latter relation is linked to the scenarios with flat extra dimension considered in this paper (for which \eqref{decadix-bis} holds true). \\

Therefore, one obtains from \eqref{decadix-bis} and \eqref{kappagbr} a lower bound on the inverse size of the extra dimension $\mu$, namely
\begin{equation}
\mu\gtrsim\mathcal{O}(10^{-6}-10^{-3})M_P\label{sizemiou},
\end{equation}
which, still assuming $M_P\sim\mathcal{O}(10^{19})$ GeV, gives $\mu\gtrsim\mathcal{O}(10^{13}-10^{16})$ GeV. \\
From this, one may conclude that if the notion of dispersion relation is meaningful in the present noncommutative context, then present GRB data on light speed variations definitely favor scenarios with {\it{very small}} flat extra dimension for $\kappa$-Poincar\'e invariant gauge theories. Note that the K-K spectrum is expected to be heavy with typical mass proportional to $\mu$.\\

At this stage, one important comment must be pointed out.\\

First, we note that our kinetic operator is, up to some (here) unessential factor, very similar to the one used in \cite{mercati}. In this recent work, a $\kappa$-Poincar\'e invariant free scalar field theory on $\mathcal{M}^4_\kappa$ is used to model a noncommutative notion of light-cone through a reasonable noncommutative extension of the Pauli-Jordan-Schwinger function. Recall that in standard commutative field theory, this function vanishes 
outside the light-cone and encodes, up to technical subtleties, the (micro)causality condition. Hence, the use of a reasonable noncommutative Pauli-Jordan-Schwinger function, as done in \cite{mercati}, actually represents a possible way to get access to some salient features of causal structures of a noncommutative space-time such as the $\kappa$-Minkowski space, once an action for a noncommutative field theory is available. Interestingly enough, the main conclusion of \cite{mercati} is that analysis based on dispersion relations confronted to GRB data cannot constraint noncommutative field theories on $\kappa$-Minkowski space. In particular, it is found that the above noncommutative light-cone is blurry, so that the notion of dispersion relation underlying most of the above analysis should be carefully reconsidered. Besides, this effect affects the usual classical light-cone over a distance $\sim\sqrt{LL_P}$ where $L_P$ is the Planck lenght, $L_P\sim10^{-35}$ m. For a typical distance for GRB $L\sim10^9$ ly, further assuming that the source is point-like, the expected time effect would be $\Delta t\sim 10^{-14}s$, far beyond the present detection capability.\\

A somewhat abrupt adaptation of our above analysis to \cite{mercati} would simply amount to replace the Planck length by the inverse 5-d bulk Planck mass $L_\kappa\sim\kappa^{-1}$ since this is this latter mass scale which appears in the expressions related to the 4-d effective action for the zero modes. In view of \eqref{grosscale}, it can be easily realized that within our 4-d effective theory, the situation would be only improved by at most three order of magnitude, as $\mathcal{O}(10^{6})L_P\gtrsim L_\kappa$. Therefore, if the causal structure used in \cite{mercati} is physically suitable, then the GRB bound \eqref{kappagbr} is questionable. \\

Fortunately, this is not the end of the story. Causality in noncommutative geometry is a delicate story.  It turns out that the notion of noncommutative causality is not unique, a fact which already temperates the above negative conclusion. For a review on actual approaches to noncommutative causality, see e.g. \cite{besnard} and references therein. \\
Among these approaches, Lorentzian noncommutative geometry \cite{petitnicolas} has been set up as an attempt to adapt to Lorentzian signature the noncommutative geometry which is usually developed in Euclidean signature. Lorentzian noncommutative geometry involves a natural notion of causality rooted to the Dirac operator playing in some sense the role of a metric and underlies the notion of noncommutative metric geometry{\footnote{In an Euclidean set-up, the Dirac operator is the building block of the spectral distance generalizing the geodesic distance. For explicit constructions on various noncommutative spaces, see \cite{spectrodist}. }}. This noncommutative causality, which coincides to the usual one at the commutative limit, has been applied for almost-commutative manifolds \cite{eckstein} and for Moyal plane \cite{francwall}, giving in that latter case the explicit characterization of the relevant causal structure and by the way closing a controversy of the physics literature about the existence of some causality on the Moyal plane at the Planck scale. \\
In this respect, the confrontation of the notion of light-cone used in \cite{mercati} to the causal cone stemming from the above noncommutative causality deserves to be carried out in order to evaluate if the negative conclusion of \cite{mercati} still survives or leads to a less pessimistic situation. Such a study has been undertaken \cite{francwall2}.\\

\section{Conclusion.}\label{section4}

We have shown that $\kappa$-Poincar\'e invariant gauge theories on $\kappa$-Minkowski space with physically suitable commutative limit must be 5-d. Only for this special value, noncommutative (quantum) gauge and $\kappa$-Poincar\'e invariances can be reconciled through the use of twisted connections and curvatures pertaining to a unique differential calculus, which singles out a unique family of twisted derivations of the algebra of the deformed translations. This result appears as an interesting physical prediction, giving as a byproduct a rationale based on symmetries for the appearance of one spatial extra dimension. \\
A BRST symmetry related to the noncommutative gauge invariance has been characterized through the definition of a nilpotent Slavnov operation $s$ and used to construct a (5-d) gauge-fixed action whose commutative limit reduces to the usual (5-d) gauge-fixed QED action in the Lorentz gauge.\\

In order to make contact with the 4-d world, we have explored and discussed some generic properties of the 4-d effective theories which would be obtained within the type of standard scenarios with (compactification of) flat extra dimension \cite{UED}, focusing mainly on the zero-modes part of the action. Here, the deformation parameter $\kappa$ can be naturally interpreted as the 5-d bulk Planck mass. \\
Consistency of these scenarios with the recent bounds from collider physics experiments \cite{bounds-hep} on the size of the compact extra dimension requires $\kappa\gtrsim\mathcal{O}(10^{13})\ \text{GeV}$ which would imply a strong suppression of the effects stemming from the underlying noncommutative structure (i.e. $\kappa$-depending) in present and (near) future collider experiments, as expected.\\

The invariances of the bulk gauge action together with the related differential calculus mentioned above give rise finally to a deformed dispersion relation in the 4-d effective theory for the zero modes, identified to photons. Assuming, as done in most of the literature, that the (usual) notion of dispersion relation is still meaningful in a noncommutative context (or at least for the $\kappa$-Minkowski space-time), the confrontation of data of GRB photons to the deformed dispersion relation \cite{ESA1}--\cite{ESA3} improves the lower bound on $\kappa$ to $\kappa\gtrsim\mathcal{O}(10^{17}-10^{18})$ GeV. Such values would favor the scenarios with a {\it{very small}} flat extra dimension for $\kappa$-Poincar\'e invariant gauge theories.\\
We have finally discussed the robustness of this latter bound if the usual notion of dispersion relation must be revised. This may be the case if the ``noncommutative'' light-cone defining the relevant causal structure deviates from the classical one, as shown recently in \cite{mercati}. Drawing a more firm conclusion requires at least to verify if the conclusions \cite{mercati} still holds true, assuming a different noncommutative causal structure.\\

\noindent{\bf{Acknowledgements:}} Discussions at various stages of this work with F. Lizzi, A. Martin, P. Martinetti, A. de Rossi, A. Sitarz and A. Wallet are gratefully acknowledged. We would like to acknowledge the contribution of the COST Action CA18108.
\appendix
\section{$\kappa$-Poincar\'e algebra and deformed translations.}\label{apendixA}

We will work in the bicrossproduct basis \cite{majid-ruegg}. We denote respectively by $\Delta:\mathcal{P}^d_\kappa\otimes\mathcal{P}^d_\kappa\to\mathcal{P}^d_\kappa$, $\epsilon:\mathcal{P}^d_\kappa\to\mathbb{C}$ and $\bf{S}:\mathcal{P}^d_\kappa\to\mathcal{P}^d_\kappa$, the coproduct, counit and antipode, thus equipping $\mathcal{P}^d_\kappa$ with a Hopf algebra structure. $\mathcal{P}^d_\kappa$ can be conveniently described from the elements $(P_i, N_i,M_i, \mathcal{E},\mathcal{E}^{-1})$, $i=1, 2, \hdots, d-1$, respectively denoting the momenta, the boosts, the rotations and $\mathcal{E}:=e^{-P_0/\kappa}$ satisfying the Lie algebra
\begin{equation}
[M_i,M_j]= i\epsilon_{ij}^{\hspace{5pt}k}M_k,\ [M_i,N_j]=i\epsilon_{ij}^{\hspace{5pt}k}N_k,\ [N_i,N_j]=-i\epsilon_{ij}^{\hspace{5pt}k}M_k\label{poinc1}, 
\end{equation}
\begin{equation}
[M_i,P_j]= i\epsilon_{ij}^{\hspace{5pt}k}P_k,\ [P_i,\mathcal{E}]=[M_i,\mathcal{E}]=0,\ [N_i,\mathcal{E}]=\frac{i}{\kappa}P_i\mathcal{E}\label{poinc2},
\end{equation}
\begin{equation}
[N_i,P_j]=-\frac{i}{2}\delta_{ij}\left(\kappa(1-\mathcal{E}^{2})+\frac{1}{\kappa}\vec{P}^2\right)+\frac{i}{\kappa}P_iP_j\label{poinc3}.
\end{equation}
The Hopf algebra structure is defined by
\begin{align}
\Delta P_0&=P_0\otimes\bbone+\bbone\otimes P_0,\ \Delta P_i=P_i\otimes\bbone+\mathcal{E}\otimes P_i,\label{hopf1}\\
\Delta \mathcal{E}&=\mathcal{E}\otimes\mathcal{E},\ \Delta M_i=M_i\otimes\bbone+\bbone\otimes M_i,\label{hopf1bis}\\
\Delta N_i&=N_i\otimes \bbone+\mathcal{E}\otimes N_i-\frac{1}{\kappa}\epsilon_{i}^{\hspace{2pt}jk}P_j\otimes M_k,\label{hopf2}\\
\epsilon(P_0)&=\epsilon(P_i)=\epsilon(M_i)=\epsilon(N_i)=0,\  \epsilon(\mathcal{E})=1\label{hopf3},\\
{\bf{S}}(P_0)&=-P_0,\ {\bf{S}}(\mathcal{E})=\mathcal{E}^{-1},\  {\bf{S}}(P_i)=-\mathcal{E}^{-1}P_i,\  {\bf{S}}(M_i)=-M_i,\label{hopf4}\\
{\bf{S}}(M_i)&=-M_i,\ {\bf{S}}(N_i)=-\mathcal{E}^{-1}(N_i-\frac{1}{\kappa}\epsilon_{i}^{\hspace{2pt}jk}P_jM_k)\label{hopf4bis}.
\end{align}
The $\kappa$-Minkowski space $\mathcal{M}_\kappa^d$ can be described as the dual of the Hopf subalgebra $\mathcal{T}_\kappa^d$ which is generated by $P_\mu$, $\mathcal{E}$. It is called the deformed translation algebra. It becomes a $^*$-Hopf algebra through: $P_\mu^\dag=P_\mu$, $\mathcal{E}^\dag=\mathcal{E}$. Then, the extension of the above duality to a duality between $^*$-algebras, which achieves the compatibility among the involutions, leads to
\begin{equation}
(t\triangleright f)^\dag={\bf{S}}(t)^\dag\triangleright f^\dag,\label{pairing-involution}
\end{equation}
for any $t$ in $\mathcal{T}_\kappa^d$ and any $f\in\mathcal{M}^d_\kappa$. This, combined with \eqref{hopf4} yields
\begin{equation}
(P_0\triangleright f)^\dag=-P_0\triangleright(f^\dag),\ (P_i\triangleright f)^\dag=-\mathcal{E}^{-1}P_i\triangleright(f^\dag),\ (\mathcal{E}\triangleright f)^\dag=\mathcal{E}^{-1}\triangleright(f^\dag)\label{dag-hopfoperat}.
\end{equation}
Note that \eqref{hopf1} implies that $P_i$'s act as twisted derivations on $\mathcal{M}_\kappa^d$ while $P_0$ remains untwisted. Indeed, one readily obtains 
\begin{align}
P_i\triangleright(f\star g)&=(P_i\triangleright f)\star g+(\mathcal{E}\triangleright f)\star (P_i\triangleright g)\label{deriv-twist1},\\
P_0\triangleright(f\star g)&=(P_0\triangleright f)\star g+f\star(P_0\triangleright  g )\label{deriv-twist2}.
\end{align}
for any $f,g\in\mathcal{M}_\kappa^d$. Note also that $\mathcal{E}$ is not a derivation of $\mathcal{M}_\kappa^d$. Indeed, the first of \eqref{hopf1bis} implies
\begin{equation}
\mathcal{E}\triangleright(f\star g)=(\mathcal{E}\triangleright f)\star(\mathcal{E}\triangleright g).\label{relation-calE}
\end{equation}
Finally, we recall the action of $\mathcal{T}_\kappa^d$ on $\mathcal{M}_\kappa^d$. It is given by 
\begin{equation}
(\mathcal{E}\triangleright f)(x)=f(x_0+\frac{i}{\kappa},\vec{x})\label{left-module0},\\
\end{equation}
\begin{equation}
(P_\mu\triangleright f)(x)=-i(\partial_\mu f)(x)\label{left-module1}.\\
\end{equation}


\end{document}